\documentstyle[psfig]{mn} 
\title{NGC~6404 and NGC~6583: two neglected intermediate-age open clusters
located in the Galactic Center direction}
\author[Carraro  at al.]        
{Giovanni Carraro$^{1,2,3}$, Ren\'e A. M\'endez$^1$, and Edgardo Costa$^1$
\thanks{email: 
gcarraro (GC), rmendez (RAM), costa (EC) @das.uchile cl}\\ 
$^1$Departamento de Astronom\'ia, Universidad de Chile, 
Casilla 36-D, Santiago, Chile\\
$^2$Astronomy Department, Yale University, 
P.O. Box 208101, New Haven, CT 06520-8101 , USA\\
$^3$Dipartimento di Astronomia, Universit\`a di Padova,
Vicolo Osservatorio 2, I-35122, Padova, Italy\\ 
 } 
 
\date{\it Submitted: August 2004} 
\pubyear{2005} 
\begin{document} 
\maketitle 
\title{The open clusters NGC~6404 and NGC~6583} 
 
\begin{abstract} 
We report on $VI$ CCD photometry of two fields centered  
in the region of the
open clusters NGC~6404 and NGC~6583 down to $V=22.0$. 
These clusters have never been studied insofar, and we provide
for the first time estimates of their fundamental parameters,
namely, radial extent, age, distance and reddening.
We find that NGC~6404 radius is 2.0 arcmin, as previously
proposed, while NGC~6583 radius is 1.0 arcmin, 
significantly lower than  previous estimates.\\
Both clusters turn out to be of intermediate age
(0.5-1.0 Gyr old), and located inside the solar ring, 
at a Galactocentric distance of about 6.5 kpc. 
These results make these objects very interesting
targets for spectroscopic follow-up to measure their metallicity. 
In fact they might allow us
to enlarge by more than 1 kpc the baseline of the radial abundance gradient
in the Galactic disk toward the Galactic Center direction. This baseline
is currently rather narrow especially for clusters of this age.
\end{abstract} 
 
 
\begin{figure*} 
\centerline{\psfig{file=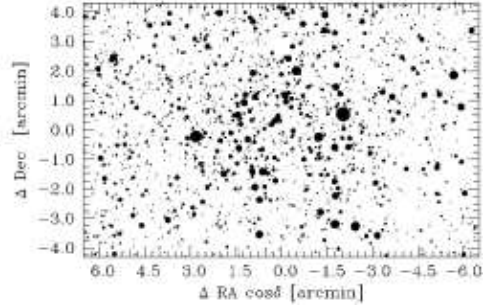}} 
\caption{A finding chart with the observed area in the region
of the open cluster NGC~6404. The sizes of the dot are proportional
to the magnitudes of the stars. North is up, east on the left, 
and the field is centered at the cluster nominal center (see Table~1)}
\label{mappa} 
\end{figure*} 
 
\begin{figure*} 
\centerline{\psfig{file=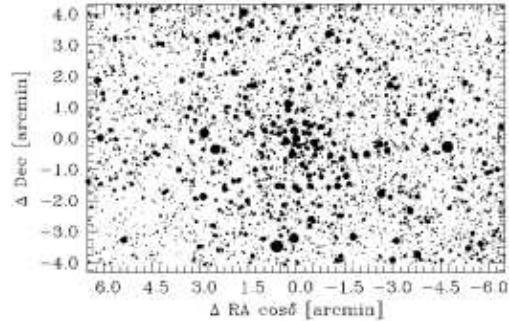}} 
\caption{A finding chart with the observed area in the region
of the open cluster NGC~6583. The sizes of the dot are proportional
to the magnitudes of the stars. North is up, east on the left, and 
the field is centered at the cluster nominal center (see Table~1)}
\label{mappa} 
\end{figure*}

\section{Introduction} 
Intermediate age and old open clusters (older than half a Gyr)
are widely used
to probe the chemical evolution of the Galactic disk 
(Friel \& Janes 1993, Carraro \& Chiosi 1994,
Carraro et al. 1998, Friel et al. 2002, Carraro et al. 2004), 
since they cover the entire life of the disk and are evenly distributed 
across the disk itself.\\
With these objects it is possible to derive the
age-metallicity  relationship and  the present and past
radial abundance gradients in the Galactic disk: these relations
are routinely used to constrain Galactic chemical evolution models (Tosi 1996).\\
One of the major limitation of the samples commonly in use is the
range in Galactocentric distances: a few clusters are known
to be located beyond 12 kpc from the Galactic center, and
none is currently known to lie closer than 7.5 kpc from the Galactic 
center (see Friel et al. 2002, Fig.~3).
This is basically due to selection effects; star clusters inside the solar
ring do not survive for enough time due to encounters with molecular clouds
and in general the higher density environment (Wielen 1971). 
On the other hand, 
in the antic-enter direction we expect quite a few clusters due to the low efficiency
of clusters formation in the Galaxy periphery.\\
\noindent
In an effort to enlarge the distance baseline of intermediate age and
old open clusters, we searched for 
candidates towards the Galactic Bulge, by using criteria similar to
those adopted by Phelps et al. (1994), i.e. the presence of a number of 
similar brightness red stars. This in fact would imply the existence
of a red clump, typical of intermediate age-old open clusters.\\
This search is complementary to our survey of the open cluster remnants  
(Villanova et al. 2004) designed to seek for
old open clusters in an advanced stage of dynamical evolution,
close to their final dissolution and merging with the general
Galactic disk field.\\
In this paper we report on NGC~6404 and NGC~6583, two clusters 
located low in the Galactic plane, not very 
far from the Galactic Center direction
(see Table~1) and which fulfill our searching criteria.\\
\noindent
The layout of the paper is as follows. Sect.~2 illustrates  
the observation and reduction strategies. 
An analysis of  the geometrical
structure and star counts in the field of the two clusters
are presented in Sect.~3, whereas a discussion of
the Color-Magnitude Diagrams (CMD) is performed in Sect.~4.
Sect.~5 deals with the determination of clusters reddening, 
distance and age and,
finally, Sect.~6 summarizes our findings.

\begin{table}
\caption{Basic parameters of the observed objects.
Coordinates are for J2000.0 equinox}
\begin{tabular}{ccccc}
\hline
\hline
\multicolumn{1}{c}{Name} &
\multicolumn{1}{c}{$RA$}  &
\multicolumn{1}{c}{$DEC$}  &
\multicolumn{1}{c}{$L$} &
\multicolumn{1}{c}{$B$} \\
\hline
& {\rm $hh:mm:ss$} & {\rm $^{o}$~:~$^{\prime}$~:~$^{\prime\prime}$} & [deg] & [deg]\\
\hline
NGC~6404        & 17:39:37 & -33:14:48 & 355.66 & -1.18\\
NGC~6853        & 18:15:49 & -22:08:12 &   9.28 & -2.53\\
\hline\hline
\end{tabular}
\end{table}

\section{Observations and Data Reduction} 
 
$\hspace{0.5cm}$
CCD $VI$ observations were carried out with the eight CCDs  mosaic camera on-board
the  1. 3m  Warsaw telescope at Las Campanas Observatory (Chile), in the nights of 
July 2 to 4, 2004. 
The two clusters were centered in chip $\#$3.
With a pixel size of $0^{\prime\prime}.26$,  and a CCD size of 4096 $\times$ 2048
pixels,  
this samples a $17^\prime.7\times8^\prime.9$ field in the sky.\\
However, we trimmed the CCD and at the end 
used in this study an actual area of  $13.8^\prime.7\times8^\prime.9$\\

\noindent
The details of the observations are listed in Table~2 where the observed 
fields are 
reported together with the exposure times, the average seeing values and the 
range of air-masses during the observations. 
Fig.~1 shows the finding chart in the area of NGC~6404, and
Fig.~2 in the area  of NGC~6583. In both figures North is up, and East on the left.\\
Both the field were centered in the clusters nominal centers (Dias et al. 2002
\footnote{http://www.astro.iag.usp.br/wilton/clusters.txt}).
However, the coordinates of NGC~6583 turned out to be slightly off-set 
(about -20$^{\prime}$)
in declination, and the new coordinates are : $\alpha=18^{\rm h}~15^{\rm m}~49 {\rm s}$,
$\delta=-22^{\circ}~08^{\prime}~30^{\prime\prime}$. We shall use these new coordinates
throughout this paper.\\
\noindent
The data have been reduced with the 
IRAF\footnote{IRAF is distributed by NOAO, which are operated by AURA under 
cooperative agreement with the NSF.} 
packages CCDRED, DAOPHOT, ALLSTAR and PHOTCAL using the point spread function (PSF)
method (Stetson 1987). 
The three nights turned out to be photometric and very stable, and therefore
we derived calibration equations for all the 141 standard stars
observed during the three nights in the Landolt 
(1992)  fields SA~104-334, PG~1323-085, PG~1657+078, PG~2213+006,
PG~1633+099, SA~110-362 and   SA~92-355 (see Table~2 for details).

\begin{table} 
\fontsize{8} {10pt}\selectfont
\tabcolsep 0.10truecm 
\caption{Journal of observations of NGC~6404, NGC~6583 and standard star fields 
(July 2-4, 2004).} 
\begin{tabular}{cccccc} 
\hline 
\multicolumn{1}{c}{Field}         & 
\multicolumn{1}{c}{Filter}        & 
\multicolumn{1}{c}{Exposure time} & 
\multicolumn{1}{c}{Seeing}        &
\multicolumn{1}{c}{Airmass}       \\
 & & [sec.] & [$\prime\prime$] & \\ 
\hline 
 NGC 6404       & V &     10,300,1200   &   1.3 & 1.06-1.15 \\ 
                & I &     10,300,900    &   1.3 & 1.06-1.15 \\
\hline 
 NGC 6583       & V &     10,300,1200   &   1.2 & 1.02-1.20 \\ 
                & I &     10,300,900    &   1.2 & 1.02-1.20 \\
\hline
SA 104-334      & V &   $3 \times$200   &   1.4 & 1.24-1.26 \\ 
                & I &   $3 \times$70    &   1.4 & 1.24-1.26 \\ 
\hline
PG 1323-085     & V &   $3 \times$90    &   1.3 & 1.13-1.53 \\ 
                & I &   $3 \times$30    &   1.3 & 1.13-1.53 \\ 
\hline
PG 1657+078     & V &   $3 \times$300   &   1.5 & 1.24-2.04 \\ 
                & I &   $3 \times$100   &   1.5 & 1.24-2.04 \\ 
\hline
PG 2213+006     & V &   $3 \times$ 80   &   1.3 & 1.14-1.34 \\ 
                & I &   $3 \times$ 30   &   1.3 & 1.14-1.34 \\ 
\hline
PG 1633+099     & V &   $3 \times$120   &   1.2 & 1.33-1.50 \\ 
                & I &   $3 \times$45    &   1.2 & 1.33-1.50 \\ 
\hline
SA 110-362      & V &   $3 \times$120   &   1.2 & 1.21-1.96 \\ 
                & I &   $3 \times$30    &   1.2 & 1.21-1.96 \\ 
\hline
SA 92-355       & V &   $3 \times$120   &   1.6 & 1.15-1.18 \\ 
                & I &   $3 \times$50    &   1.6 & 1.15-1.18 \\ 
\hline
\hline
\end{tabular}
\end{table}

\noindent
The calibration equations turned out of be of the form:\\

\noindent
$ v = V + v_1 + v_2 * X + v_3~(V-I)$ \\
$ i = I + i_1 + i_2 * X + i_3~(V-I)$ ,\\

\noindent

\begin{table} 
\tabcolsep 0.3truecm
\caption {Coefficients of the calibration equations}
\begin{tabular}{ccc}
\hline
$v_1 = 2.029 \pm 0.005$ & $v_2 =  0.15 \pm 0.02$ & $v_3 = -0.022 \pm 0.005$ \\
$i_1 = 2.002 \pm 0.005$ & $i_2 =  0.07 \pm 0.02$ & $i_3 =  0.072 \pm 0.005$ \\
\hline
\end{tabular}
\end{table}

\noindent
where $VI$ are standard magnitudes, $vi$ are the instrumental ones and  $X$ is 
the airmass; all the coefficient values are reported in Table~3.
The standard 
stars in these fields provide a very good color coverage.
The final {\it r.m.s.} of the calibration are 0.034 and 0.033 for the V and I filter, respectively.

\noindent
Photometric errors have been estimated following Patat \& Carraro (2001).

It turns out that stars brighter than  
$V \approx 20$ mag have  
internal (ALLSTAR output) photometric errors lower 
than 0.10~mag in magnitude and lower than 0.18~mag in colour, as one can readily see
by inspecting Fig.~3. There the trend of errors in colour and magnitude
are reported against the V mag., while in the insert we show 
the mean errors as a function of the magnitude.\\
\noindent
The final photometric data (coordinates,
V and I magnitudes and errors)  
consist of 24,295 stars in NGC~6404 and 26,086 stars in NGC~6583, and
are made 
available in electronic form at the  
WEBDA\footnote{http://obswww.unige.ch/webda/navigation.html} site
maintained by J.-C. Mermilliod.\\

\begin{figure}
\centering
\centerline{\psfig{file=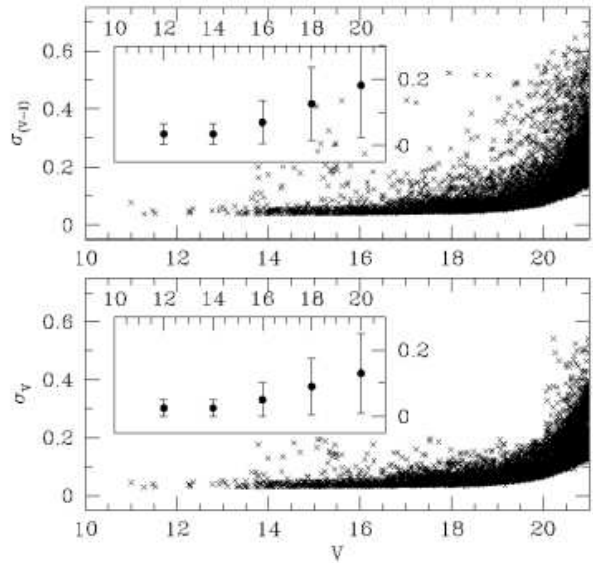,width=\columnwidth}} 
\caption{Trend of  photometric errors in V and (V-I)
as a function of V magnitude.}
\end{figure}

\begin{figure} 
\centerline{\psfig{file=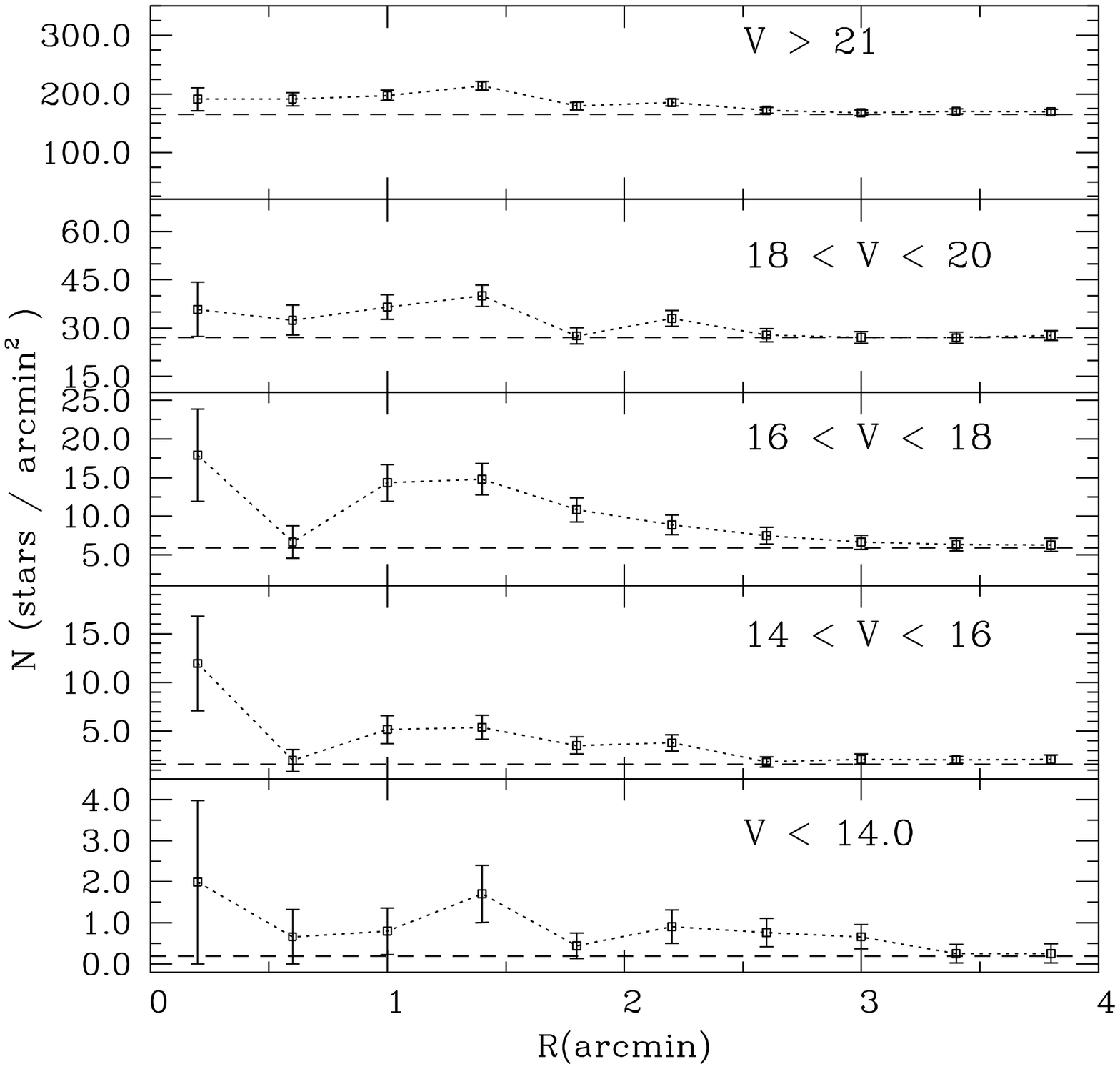,width=\columnwidth}} 
\caption{Star counts in the area of 
NGC~6404 as a function of  radius and magnitude. The dashed lines represent
the level of the control field counts estimated in the surroundings
of the cluster in that magnitude range.}
\end{figure} 

\begin{figure} 
\centerline{\psfig{file=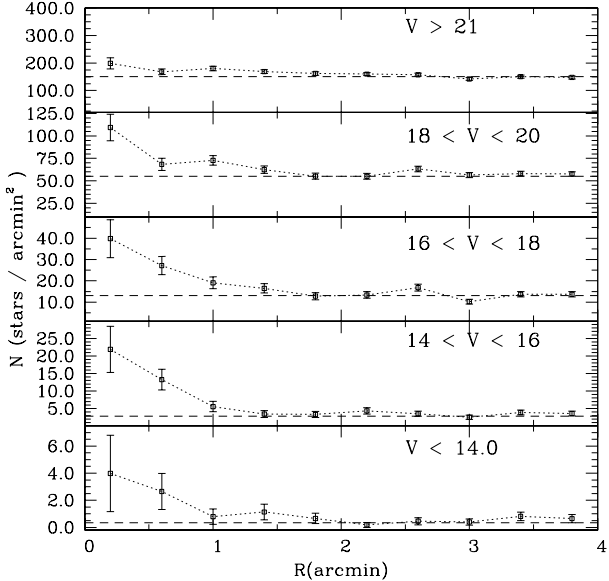,width=\columnwidth}}
\caption{Star counts in the area of 
NGC~6583 as a function of radius and magnitude. The dashed lines represent
the level of the control field counts estimated in the surroundings
of the cluster in that magnitude range .}
\end{figure} 

\section{Star counts and clusters size} 
Dias et al. (2002) report preliminary estimates
of NGC~6404 and NGC~6583 diameters amounting to 5 arcmin. 
By inspecting Fig~1 and 2 we can recognize that Dias et al. estimate
is surely a reasonable one for NGC~6404, which is a loose
open cluster, but it seems to be too large
for NGC~6583, which on the contrary appears more concentrated.\\
Since our photometry covers entirely the clusters area and part
of the surroundings, we performed star counts to obtain
an improved estimate of the clusters size.\\
We derived the surface stellar density by performing star counts
in concentric rings around the clusters nominal centers (see Table~1)
and then dividing by their
respective surfaces. Poisson errors have also been derived and normalized
to the corresponding surface. Poisson errors in the field star counts
turned out to be very small, and therefore we are not going to show them.

\subsection{NGC~6404}
The final radial density profile for NGC~6404 is shown in Fig.~4
as a function of V magnitude.
The contribution of Galactic disk field has been estimated by considering
all the stars in the corresponding
magnitude bin, located outside 4.0 arcmin from the cluster center, 
and by normalizing counts over the adopted area.\\ 
The cluster seems to be populated  by stars of magnitude in the range
$12 \leq V \leq 18$, where it clearly emerges from the background,
and then it starts to be well mixed with the field.
In this magnitude range the radius is not larger than 2 arcmin,
and the cluster exhibits a significant under-density of stars (at the level of the field)
at about half an arcmin from the nominal center. This is compatible with
the loose nature of NGC~6404 (see also Fig.~1).\\
\noindent
In conclusion,  we are going to adopt the  value of 2 arcmin as NGC~6404 
radius throughout this paper. This estimate is in good agreement with
the value reported by Dias et al. (2002).

\begin{figure*} 
\centerline{\psfig{file=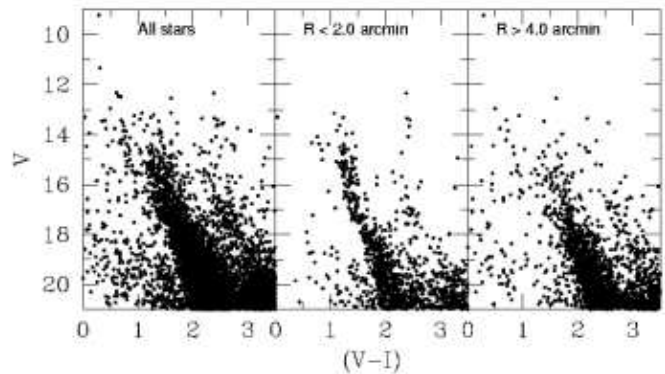}} 
\caption{$V$ vs $(V-I)$ CMDs
of NGC~6404 as a function of  radius from the cluster center.}
\end{figure*} 

\subsection{NGC~6583}
The final radial density profile for NGC~6583 is shown in Fig.~5
as a function of V magnitude.
Also in this case the contribution of Galactic disk field has been estimated by considering
all the stars outside 4.0 arcmin from the cluster center in the same way as for NGC~6404.
Unlike NGC~6404, NGC~6503 is a compact cluster, which clearly emerges
above the background down to $V \approx 20$.
The cluster radius turns out of be around 1 arcmin. 
Within this radius,
the cluster exhibits a significant over-density of stars. Outside, 
the counts level off to the field star counts value.\\
\noindent
We thus adopt the  value of 1 arcmin as NGC~6503 
radius throughout this paper. This estimate is a factor of two  smaller than that
reported by Dias et al. (2002).

\begin{figure*} 
\centerline{\psfig{file=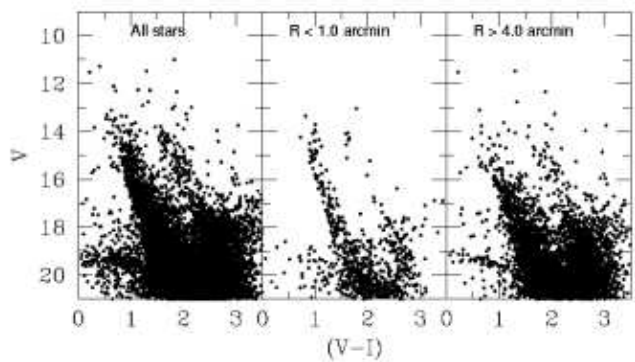}} 
\caption{$V$ vs $(V-I)$ CMD
of NGC~6583 as a function of radius from the cluster center.}
\end{figure*} 

\section{The Colour-Magnitude Diagrams} 
In Figs. 6 and 7 we present CMDs of NGC~6404 and NGC~6583,
respectively. They are plotted
as a function of radius, in order to facilitate their 
interpretation. In fact the clusters are located quite low onto
the Galactic plane toward the Galactic Center direction, and
hence we expect quite a significant contamination from the
Galactic disk field stars located in the foreground along the 
line of sight of the clusters.\\
The cuts according to  radius are done
on the basis of the results on Sect.~3.

\subsection{NGC~6404}
The CMDs of NGC~6404 are shown in Fig.~6.
In the left panel  we plot all the
detected stars. Here the
Main Sequence (MS) extends from $V$=14.5 to $V$=21.0,
and the Galactic disk Red Giant Branch (RGB) sequence  
departs from $V$=20. The MS is very wide, and this could have
different causes, like variable reddening across
the observed area (actually we expect this is the major cause), 
 photometric errors which increase as a function of 
magnitude (see Fig.~3), and the presence of a number of binary stars.
It is very difficult to
distinguish from this CMD the presence of a cluster.
However, and interestingly,
there are a few stars in the red part of the CMD 
at $V$=13.5-14.5, $(V-I)$=2.0-2.5, which resemble
a RGB clump. \\
Much better information can be obtained by looking at the middle
and right panels in the same figure. The middle panel contains only the stars
located inside the estimated cluster radius (2 arcmin, see  Sect.~3), whereas
the right panel contains the stars located outside 4 arcmin
from the cluster center, where we estimated the contribution of the field
population to be dominant.\\ 
The following remarks can be done closely inspecting these two panels:

\begin{itemize}
\item The MS and the Turn Off Point (TO) region in the middle panel are 
much better defined, although the MS is still somewhat wide, mostly
due to field star contamination;
\item Almost all the probable RGB stars are inside the inner
region, which implies by the way that the cluster underwent some dynamical 
relaxation;
\item  Most of the stars above the TO are probably field stars, since they
lie almost all out of the cluster radius (see right panel); nevertheless some of them still
remain, and they might be blue straggler stars, quite common in clusters like this.
\end{itemize}

\noindent
The shape of the TO and a presence of some clump stars are
a clear indication of an age
in the range 0.5-1.5 Gyr, depending on the precise metal content of the cluster
(Carraro \& Chiosi 1994, Carraro et al. 1999).

\subsection{NGC~6583}
The CMDs of NGC~6583 are shown in Fig.~7, which is similar
to Fig.6.
In the left panel we plot all the
detected stars. Here the
Main Sequence (MS) extends from $V$=14.5 to $V$=21.0,
and the Galactic disk RGB sequence  
departs from $V$=190. 
Like NGC~6404, it is very difficult to
distinguish from this CMD the presence of a cluster, and we do not notice
any candidate RGB clump star.\\
Much better information can be obtained by looking at the middle
and right panels in Fig.~7. The middle panel contains only the stars
located inside the estimated cluster radius (1 arcmin, see Sect.~3), whereas
the right panel contains the stars located outside 4 arcmin
from the cluster center, where we estimate the contribution of the field
population to be dominant.\\  
The following considerations can be done:

\begin{itemize}
\item The MS and the TO region in the middle panel are 
much better defined; in particular the MS is quite narrow and
the field star contamination is almost negligible down to $V \approx 19.0$;
\item  There is  a nice almost vertical clump of stars at 
$V$=14.5, $(V-I)$=1.5, similar to the clump observed in  open clusters
like NGC~2477 (Kassis et al. 1997) or Pismis~2 (Phelps et al. 1994).
\item  Most of the stars above the TO are probably field stars, since they
lie all out of the cluster radius (see right panel).
\end{itemize}

\noindent
In particular the fine shape of the TO deserves some attention. In fact
the shape of the TO is that one typical of intermediate-age open clusters,
with a blue and red hook clearly visible, notwithstanding some field star
contamination. 
Again, the shape of the TO and the presence of a clump indicate an age
in the range 0.5-1.5 Gyr, depending on the metallicity.

\section{Cluster fundamental parameters} 
In this section we provide some estimates of the clusters
basic parameters. To achieve this, we make use of the comparison
between the stars distribution in the CMD and a set of theoretical isochrones
from the Padova group (Girardi et al. 2000).
We already have an indication of the cluster age, but we do not know
anything about the reddening, the distance, and the metallicity.
In the following analysis we adopt $R_{\odot}$=8.5 kpc for the Galactocentric
distance of the Sun, $R_V~=~3.1$ and the ratio
$\frac{E(V-I)}{E(B-V)}=1.244$ from Dean et al. (1978).
The results of the fits are shown in Fig.~8 for NGC~6404
and in Fig.~9 for NGC~6583.

\subsection{NGC~6404}
In details, in Fig.~8 we present the CMD for the stars within 2.0 arcmin
from the cluster center (see Sect.~3), and over-imposed an isochrone of 0.5 billion
years for solar (Z=0.019) metallicity. The fit is quite good both
in the TO and the evolved stars region. The fit is poor in the bottom
of the MS, where, by the way, it is not easy to distinguish the cluster
MS from the field.
We are keen to believe that the bulk of stars above the TO are most probably
field stars.\\
We achieved this results by shifting the isochrone with E$_{(V-I)}=1.15\pm0.05$
(E$_{(B-V)}=0.92$), and $(m-M)=14.75\pm0.20$ (errors by eye).\\
We also tried to over-impose a lower metal abundance
but the fit turned out to be quite poor. The same occured with higher
metallicity isochrones.\\
\noindent
Therefore, we suggest that the cluster possesses a solar metal abundance.\\
If this is the case, NGC~6404 turns out to be located 1.7 kpc from the Sun
toward the Galactic center direction. This implies a distance from the Galactic
center of 6.8 kpc and a height above 
the Galactic plane of about -40 pc. According to Friel et al. (2002, Fig.~3), 
NGC~6404 turns out to be an intermediate-age open cluster 
located 1 kpc away from the lower 
distance edge of the radial abundance gradient. \\
Therefore NGC~6404 might play an important role in defining the precise shape
of the radial abundance gradient in the inner region of the Galactic disk.

\begin{figure} 
\centerline{\psfig{file=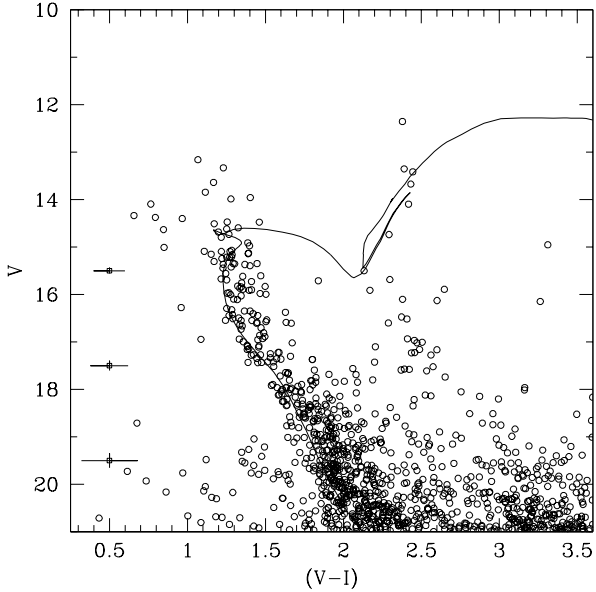,width=\columnwidth}} 
\caption{NGC~6404 data in the $V$ vs.\ $V-I$ diagram,  
as compared to Girardi et al. \ (2000) isochrone of age 
0.5 Gyr 
(solid line), for a metallicity $Z_{\odot}=0.019$. A distance 
modulus of $(m-M)_0=11.20\pm0.20$ mag, and a colour excess of E$_{(V-I)}=1.15\pm0.05$ mag, 
are derived. Errors in colour and magnitude at different magnitude levels are also
shown.} 
\end{figure} 
 
\begin{figure} 
\centerline{\psfig{file=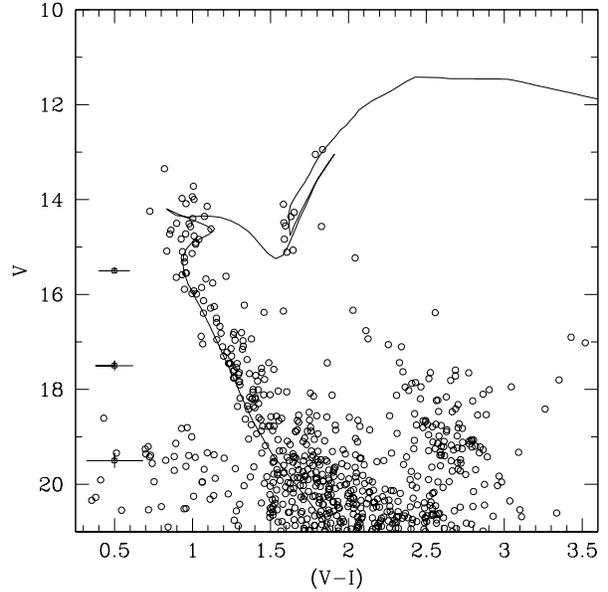,width=\columnwidth}} 
\caption{NGC~6583 data in the $V$ vs.\ $V-I$ diagram,  
as compared to Girardi et al. \ (2000) isochrone of age 
1.0 Gyr 
(solid line), for a metallicity $Z_{\odot}=0.019$. A distance 
modulus of $(m-M)_0=11.55\pm0.20$ mag, and a colour excess of E$_{(V-I)}=0.63\pm0.05$ mag, 
are derived. Errors in colour and magnitude at different magnitude levels are also
shown. } 
\end{figure}

\subsection{NGC~6583}
In Fig.~9 we present the CMD for NGC~6583 stars located within 1.0 arcmin
from the cluster center (see Sect.~3), and super-imposed an isochrone of 1.0 billion
years for a solar (Z=0.019) metallicity. The fit is quite good both along the MS,
in the TO region and in the evolved stars one. \\
We achieved this results by shifting the isochrone with E$_{(V-I)}=0.63\pm0.05$
(E$_{(B-V)}=0.51$), and $(m-M)=13.50\pm0.20$ (errors by eye).\\
Like NGC~6404, we also tried to over-impose a lower metal abundance,
but the fit turned out to be quite poor. The same occured with higher
metallicity isochrones. \\
\noindent
Therefore, we suggest that this cluster also possesses a solar metal abundance.\\
As a consequence, NGC~6503 turns out to be located 2.1 kpc from the Sun
toward the center direction. This implies a distance from the Galactic
center of 6.4 kpc and a height above 
the Galactic plane of about -90 pc. 
As NGC~6404, NGC~6583 turns out to be an intermediate-age open cluster 
located more than 1 kpc away from the lower 
distance edge of the radial abundance gradient. \\
Therefore, also NGC~6583 might play an important role in defining the precise shape
of the radial abundance gradient in the inner regions of the Galactic disk.

\begin{table*}
\caption{{}Fundamental parameters of the observed objects.}
\fontsize{8} {10pt}\selectfont
\begin{tabular}{ccccccccc}
\hline
\multicolumn{1}{c} {$Name$} &
\multicolumn{1}{c} {$Radius(arcmin)$} &
\multicolumn{1}{c} {$E(V-I)$}  &
\multicolumn{1}{c} {$E(B-V)$}  &
\multicolumn{1}{c} {$(m-M)_0$} &
\multicolumn{1}{c} {$X(kpc)$} &
\multicolumn{1}{c} {$Y(kpc)$} &
\multicolumn{1}{c} {$Z(kpc)$} &
\multicolumn{1}{c} {$Age(Gyr)$} \\
\hline
NGC~6404 & 2.0 & 1.15 & 0.92 & 11.30 & 6.80 & -0.14 & -0.04 & 0.5\\
NGC~6583 & 1.0 & 0.63 & 0.51 & 11.55 & 6.40 &  0.35 & -0.09 & 1.0\\
\hline
\end{tabular}
\end{table*}

\section{Conclusions}
We have presented the first CCD $VI$ photometric study of the 
open clusters NGC~6404 and NGC~6583. The CMDs we derive allow us to 
infer estimates of the cluster basic parameters, which
are summarized in Table~4.\\
\noindent
In detail, we find that:
 
\begin{description} 
\item $\bullet$ Both  clusters are of intermediate age;
NGC~6404 is 0.5 Gyr old, NGC~6583 1.0 Gyrs old; 
\item $\bullet$ The reddening $E_{B-V}$ turns out to be 0.92$\pm$0.05
for NGC~6404 and $0.51\pm0.05$ for NGC~6583; for both clusters solar 
metallicity isochrones provide a reasonable fit across the whole CMDs;
by the way this metal abundance is not unexpected at the clusters position
; in fact for this age range the radial abundance gradient is almost flat
around the solar metallicity;
\item $\bullet$ We place NGC~6404  and NGC~6583 
at about 1.7 and 2.1 kpc from the Sun toward 
the Galactic Center direction; 
\item $\bullet$ This way they both turn out to be 
intermediate-age open clusters located inside the solar ring, in a region 
from the Galactic Center where clusters of this age were never found insofar. 
\end{description}

\noindent
Future work should concentrate on obtaining an estimate of the cluster
metal abundance through spectra of the RGB stars. The knowledge of the
cluster metallicity, which we could not constrain very well, is of paramount
importance to better probe the trend of  metallicity across the whole
Galactic disk (Friel et al. 2002).

\section*{Acknowledgements} 
The observations presented in this paper have been carried out at 
Las Campanas Observatory (Chile).
The work of G. Carraro is supported by {\it Fundaci\'on Andes}.
R. A. M\'endez acknowledges support from the
Chilean {\sl Centro de Astrof\'\i sica} FONDAP No. 15010003.
This study made use of Simbad and WEBDA databases.

\end{document}